S. M. Riaz, M. C. Lucas-Estañ, B. Coll-Perales and J. Gozalvez, "Predictive Configured Grant Scheduling for Deterministic Wireless Communications", *Proc. of the 2025 IEEE 101st Vehicular Technology Conference (VTC2025-Spring)*, Oslo, Norway, 2025, pp. 1-5. DOI: 10.1109/VTC2025-Spring65109.2025.11174458.

# Predictive Configured Grant Scheduling for Deterministic Wireless Communications

Syed Morsleen Riaz, M.Carmen Lucas-Estañ, Baldomero Coll-Perales, Javier Gozalvez
*Uwicore laboratory, Universidad Miguel Hernández de Elche, Elche (Alicante), Spain*
sriaz@umh.es, m.lucas@umh.es, bcoll@umh.es, j.gozalvez@umh.es

***Abstract*— Future wireless networks must enhance their capacity to sustain deterministic service levels and support emerging time-sensitive services in key verticals. The ability to guarantee bounded latencies heavily depends on efficient radio resource management. Configured Grant (CG) scheduling can reduce latency by pre-allocating resources, but its effectiveness and efficiency decrease under variable traffic patterns. This study presents a novel predictive CG scheduling scheme that pre-allocates resources based on traffic predictions while accounting for prediction inaccuracies. By considering these inaccuracies, the scheme significantly improves the ability to meet bounded latency requirements, which are essential for supporting deterministic service levels. Our evaluations show that the proposed scheme significantly enhances the capacity to support deterministic service levels while improving resource utilization, even in scenarios with variable and mixed traffic flows with diverse requirements.**

***Keywords—scheduling, configured grant, predictive, deterministic, time-sensitive, 5G, 6G.***

## I. INTRODUCTION

5G networks were designed to support ultra-reliable low-latency communications (URLLC) and new verticals such as smart mobility and manufacturing. However, the increasing digitalization and automation in these verticals introduce new challenges to support deterministic services with stringent bounded latency requirements. To meet these demands, future networks must support deterministic communications while accommodating mixed traffic flows with varying characteristics and diverse QoS (Quality of Service) requirements. The ability to sustain deterministic service levels relies heavily on efficient and effective radio resource management, including advanced scheduling mechanisms that can anticipate demands and dynamically schedule transmissions in mixed traffic environments.

5G and beyond can reduce latencies through semi-static scheduling schemes, which include Configured Grant (CG) for uplink (UL) transmissions and Semi-Persistent Scheduling (SPS) for downlink (DL) transmissions. Semi-static scheduling eliminates the need to send a Scheduling Request (SR) and/or wait for a Scheduling Grant before transmitting a data packet, instead pre-allocating resources to nodes so they can immediately transmit packets upon generation. Semi-static scheduling has proven highly effective in ensuring low transmission latencies for periodic traffic with fixed packet sizes. However, its effectiveness and efficiency decrease in scenarios where packet sizes vary, message periodicity fluctuates or does not align with the periodicity of resource allocation [1], or when multiple traffic flows with different periodicities coexist [2]. To address these inefficiencies, current approaches explore predictive scheduling mechanisms that anticipate traffic demands. Studies such as [3] propose semi-static scheduling schemes that pre-allocate or reserve resources based on traffic predictions. However, such pre-allocations are prone to inefficiencies due to prediction inaccuracies and the inherent stochasticity of wireless systems. These inefficiencies can negatively impact overall system capacity and compromise the ability to guarantee bounded latency requirements for deterministic services [3]. In this context, this work proposes a novel predictive Configured Grant (CG) scheduling scheme that accounts for inaccuracies in traffic predictions when pre-allocating radio resources based on anticipated traffic demands. The proposed CG scheduling scheme is designed to support deterministic service levels in scenarios with mixed traffic flows and diverse QoS requirements. Rather than minimizing transmission latency, the scheme aims to maximize the percentage of transmissions that meet their bounded latency requirements. By leveraging predicted traffic information along with potential prediction inaccuracies, the scheme pre-allocates resources with a high likelihood of meeting bounded latency requirements. Our evaluation demonstrates that the proposed predictive CG scheduling scheme significantly improves both the capacity to support deterministic service levels and resource utilization efficiency under mixed traffic flows with diverse requirements compared to a standard CG scheduling.

This work has been partially funded by the European Commission Horizon Europe SNS JU 6G-SHINE (GA 101095738) project, and by MCIN/AEI/10.13039/501100011033 (PID2023-150308OB-I00), and the "European Union NextGenerationEU/PRTR" (TED2021-130436B-I00), by Generalitat Valenciana and UMH's Vicerrectorado de Investigación grants.

## II. STATE OF THE ART

The 3GPP standard [4] defines two types of Configured Grant (CG). In *Type 1*, the uplink grant is configured through Radio Resource Control (RRC) signaling. In *Type 2*, RRC signaling only defines the grant's periodicity, while the uplink grant is signaled, activated, and deactivated using the PDCCH control channel, similar to Semi-Persistent Scheduling (SPS). CG Type 2 and SPS offer greater flexibility, enabling the dynamic adaptation of the configured grant based on changes in network or traffic conditions. Several studies have explored methods to enhance the adaptability of semi-static scheduling, improving its effectiveness while optimizing radio resource utilization. For example, [5] uses offline and online learning to dynamically adjust resource allocations in CG scheduling. The proposal continuously monitors parameters such as node buffer status and wasted resources to dynamically determine an optimal allocation strategy that minimizes the cumulative buffer status of UEs while ensuring a fair distribution of resources. The proposal reduces latency under stable traffic and channel conditions but faces challenges in highly dynamic environments. In [6], the authors propose a learning-based approach to periodically determine and adjust the allocation of radio resources for CG scheduling. The study focuses on using CG with shared resources for uplink traffic in a massive Machine-Type Communication (mMTC) scenario with heterogeneous MTC devices. The proposed scheme organizes nodes into priority-based groups based on their requirements and dynamically adjusts resource allocations according to the estimated traffic arrival rate and priority level. Similarly, [7] explores a scenario where mMTC and eMBB nodes share radio resources. mMTC nodes sense transmissions from eMBB nodes to detect patterns and use reinforcement learning to autonomously select radio resources in a grant-free scheduling framework, thereby avoiding collisions with eMBB transmissions. Despite the gains achieved, shared resources can compromise the ability to meet bounded latency deadlines for deterministic services, particularly as traffic demand increases.

Several studies propose leveraging predictive techniques to forecast traffic demand and proactively schedule resource allocations. For instance, [8] introduces a scheduling scheme that assigns radio resources within a scheduling window of *m* consecutive slots to the highest-priority nodes. A node's priority is determined based on factors such as the amount of buffered data, the predicted data generation in the next scheduling period of *m* slots, and the predictive average data rate for the upcoming *m* slots. In [9], the authors propose using proactive grants to allocate resources based on predictions of the data different nodes are expected to generate. A first approach predicts data sizes and allocates the necessary radio resources for each user in every slot. While this minimizes latency, it leads to over-reserving radio resources that may go unused. A second approach enhances efficiency by predicting both data sizes and traffic generation times. However, inaccuracies in traffic predictions can cause some packets to experience increased latency. In this context, this study progresses the state-of-the art with a novel predictive CG scheduling scheme that pre-allocates resources based on traffic predictions and prediction inaccuracies. By considering these inaccuracies, the scheme improves the ability to meet bounded latency requirements, and hence support deterministic services.

## III. Predictive Configured Grant Scheduling

We consider a scenario with $N$ nodes, where each node $i$ ($i$=0, 1,…,$N$-1) generates data with an inter-packet generation time characterized by a mean value $p_i$ and a standard deviation $\sigma_i^t$. The packet size is also characterized by a mean value $s_i$ and a standard deviation $\sigma_i^s$. Each packet has a latency requirement $L_i$. Without loss of generality, we consider $p_i = p$ ms for all $N$ nodes. The proposed predictive CG scheduler determines the radio resources to be allocated for all nodes within a scheduling interval of $p$ ms, and scheduling decisions can be updated at every scheduling interval $p$. Resources are proactively pre-allocated to maximize the likelihood of satisfying the requirements of each transmission. To achieve this, a predictor forecasts the generation time $\hat{t}_i$ and size $\hat{s}_i$ of the next packet $pkt_i$ for each node within the upcoming scheduling interval $p$. The scheduler then calculates the number of resources required for transmitting each packet $pkt_i$, considering not only the predicted size $\hat{s}_i$ but also a margin $\Delta s_i$ to account for prediction inaccuracies; $\Delta s_i$ can be set equal to the standard deviation $\sigma_i^s$ or other value related to the prediction inaccuracy. We assume a NR radio interface where a radio resource is defined by a Resource Block (RB) in the frequency domain and a slot in the time domain. The scheduler estimates the number of RBs needed for $pkt_i$ as $R_i = f(\hat{s}_i+\Delta s_i, mcs_i)$, where $f(\cdot)$ is a function that determines the number of RBs required to transmit a packet of size $\hat{s}_i+\Delta s_i$ using a Modulation and Coding Scheme (MCS) $mcs_i$ following [4]. The proposed scheduler seeks to allocate $R_i$ RBs for $pkt_i$ within the allocation window $w_i=w(t_i^{ini}, t_i^{end})$, where $t_i^{ini} = \hat{t}_i+\Delta t_i$ and $t_i^{end} = \hat{t}_i-\Delta t_i+L_i$, $\hat{t}_i$ is the predicted generation time of $pkt_i$ and $\Delta t_i$ is a time margin to account for prediction inaccuracies of the generation time. The $w_i$ ensures that a packet generated between $\hat{t}_i-\Delta t_i$ and $\hat{t}_i+\Delta t_i$ can be transmitted within its latency requirement $L_i$, as shown in Fig. 1. The scheduler identifies $w_i$ for all packets $pkt_i$ within the scheduling interval $p$, and searches for a scheduling solution that meets the latency requirements of the maximum possible number of packets. We assume a mixed traffic flow scenario where packets may have different latency requirements.

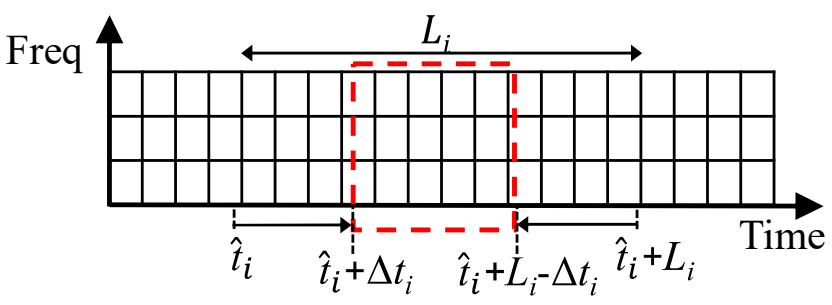

Fig. 1. Allocation window $w_i$ for packet $pkt_i$.

The operation of the proposed predictive CG scheduling scheme is presented in Algorithm I. The scheme defines the set $\Phi$, which is initialized with all the packets $pkt_i$ scheduled for the next scheduling interval $p$. The scheduler starts processing packets with the shortest latency requirements (line 2 in Algorithm I). For each packet $pkt_i$, the scheduler first searches for available RBs within its allocation window $w_i$ that do not overlap with the allocation window $w_j$ of other packets $pkt_j$ ($\forall j\neq i$) (line 3 of Algorithm I). We define this non-overlapping area of $w_i$ as $w_i^{no} = w(t_i^{no\text{-}i}, t_i^{no\text{-}e})$. If the number of available RBs $R_i^{no}$ within $w_i^{no}$ is smaller than $R_i$ ($R_i = f(\hat{s}_i+\Delta s_i, mcs_i)$), then $pkt_i$ is added to an auxiliar set $\Theta$ as a pending packet awaiting RB allocation. Conversely, if $R_i^{no} \geq R_i$, the scheduler allocates $R_i$ RBs to $pkt_i$ within $w_i^{no}$ (lines 4-9 of *Search_in_non-overlapping_area*). If multiple allocation options exist, the scheduler follows the policy outlined in *allocate_RB*. In particular, if $w_i$ does not overlap with any other $w_j$ ($\forall j\neq i$), the scheduler allocates RBs for $pkt_i$ at the center of $w_i$. This increases robustness against inaccuracies in the estimated generation time of $pkt_i$. If $w_i$ partially overlaps with the allocation window $w_j$ of other packets, the scheduler prioritizes allocating resources from the boundary of $w_i^{no}$ that is furthest from the respective limits $t_i^{ini}$ or $t_i^{end}$ of $w_i$. This ensures that packets generated with a deviation larger than $\Delta t_i$ from the estimated time $\hat{t}_i$ can still meet their latency requirements. Once $pkt_i$ receives RBs, it is removed from $\Phi$ (line 10 in *Search_in_non-overlapping_area*), and $w_i$ is no longer considered when allocating RBs for other packets $pkt_j$ ($j\neq i$). After completing the allocation process for all packets in $\Phi$, the process in lines 2-13 in *Search_in_non-overlapping_area* is repeated for the remaining packets in $\Theta$. We should note that a packet $pkt_j$ that previously had $R_i^{no}<R_i$ might now have $R_i^{no}\geq R_i$ after removing from $\Phi$ the packets $pkt_j$ that successfully received RBs. This iterative process continues until no further packets can be allocated the required $R_i$ resources within non-overlapping areas.

For packets that cannot be assigned the required resources $R_i$ within non-overlapping areas, the scheduler follows *Search_in_overlapping_area* (line 5 in Algorithm I). The process begins with packets that have the lowest latency requirements. For each packet $pkt_i$ remaining in $\Phi$ after

**Algorithm I**: Predictive CG Scheduling

Input: set $\Phi$ of packets to schedule, $\hat{t}_i$, $\Delta t_i$, $\hat{s}_i$, $\Delta s_i$, $L_i$ $\forall i$

1. Define for each $pkt_i$ in $\Phi$: $R_i = f(\hat{s}_i + \Delta s_i, mcs_i)$, $w_i = w(t_i^{ini}, t_i^{end})$, $t_i^{ini} = \hat{t}_i + \Delta t_i$, $t_i^{end} = \hat{t}_i + L_i - \Delta t_i$
2. Sort packets in $\Phi$ by $L_i$
3. Call $\Phi$=*Search_in_non-overlapping_area*($\Phi$)
4. If $\Phi \neq \emptyset$ (there are packets without RBs)
5.    Call *Search_in_overlapping_area*($\Phi$)
6. End procedure

**Procedure I**: *Search_in_non-overlapping_area*($\Phi$)

1. Set $\Theta = \emptyset$ ($\Theta$ is a set with pending packets)
2. Repeat
3.    $M$ = size($\Phi$)
4.    For each $pkt_i$ in $\Phi$
5.       Identify non-overlapping area $w_i^{no} = w(t_i^{no\text{-}i}, t_i^{no\text{-}e})$ between $w_i$ and $w_j$ $\forall j \neq i$ and $pkt_j$ in $\Phi$
6.       $R_i^{no}$ = available RBs within $w_i^{no}$
7.       If $R_i^{no} \geq R_i$
8.          Allocate $R_i$ RBs in $w_i^{no}$ to $pkt_i$: call *allocate_RB*
9.       Else $\rightarrow$ $pkt_i$ is included in $\Theta$
10.       Removed $pkt_i$ from $\Phi$
11.    End For
12.    $\Phi = \Theta$, $\Theta = \emptyset$
13. Until size($\Phi$) == $M$
14. Return $\Phi$ and end procedure

**Procedure II**: *allocate_RB*($pkt_i$, $w_i^{no} = w(t_i^{no\text{-}i}, t_i^{no\text{-}e})$, $R_i$, $t_i^{ini}$, $t_i^{end}$)

1. If $t_i^{no\text{-}i} == t_i^{ini}$ & $t_i^{no\text{-}e} == t_i^{end}$
2.     Allocate $R_i$ RBs in the center of $w_i^{no}$
3. Elseif $t_i^{no\text{-}i} - t_i^{ini} > t_i^{end} - t_i^{no\text{-}e}$
4.     Allocate $R_i$ RBs in $w_i^{no}$ starting from $t_i^{no\text{-}i}$
5. Else
6.     Allocate $R_i$ RBs in $w_i^{no}$ ending at $t_i^{no\text{-}e}$
7. End procedure

**Procedure III**: *Search_in_overlapping_area*($\Phi$)

1. For each $pkt_i$ in $\Phi$
2.    Repeat
3.       $R$ = available RBs within $w_i = w(t_i^{ini}, t_i^{end})$
4.       If $R \geq R_i$ $\rightarrow$ Allocate $R_i$ RBs in $w_i$ to $pkt_i$
5.       Elseif odd iteration $\rightarrow$ $t_i^{end} = t_i^{end} + \Delta t_i$
6.       If even iteration $\rightarrow$ $t_i^{ini} = t_i^{ini} - \Delta t_i$
7.    Until $pkt_i$ receives RBs or $w_i == w(\hat{t}_i - \Delta t_i, \hat{t}_i + L_i + \Delta t_i)$
8.    If $pkt_i$ has not received RBs & $R_i > f(\hat{s}_i, mcs_i)$
9.       $R_i = f(\hat{s}_i, mcs_i)$, $t_i^{ini} = \hat{t}_i + \Delta t_i$, $t_i^{end} = \hat{t}_i + L_i - \Delta t_i$
10.       Goto line 2
11. End For
12. End procedure

*Search_in_non-overlapping_area*, the scheduler checks whether there are sufficient RBs within $w_i$ to satisfy $R_i$ ($R_i = f(\hat{s}_i + \Delta s_i, mcs_i)$). If so, the scheduler allocates $R_i$ RBs within $w_i$ to $pkt_i$ (lines 3-4 in *Search_in_overlapping_area*). If there are insufficient RBs, the size of $w_i$ is iteratively augmented by $\Delta t_i$ until there are sufficient RBs available within $w_i$ to satisfy $R_i$, or $w_i$ reaches its maximum possible size $L_i + 2 \cdot \Delta t_i$ (lines 5-7 in *Search_in_overlapping_area*). The maximum size is defined so that it is still possible to meet the latency requirement $L_i$ if the packet is generated at $\hat{t}_i + \Delta t_i$ or $\hat{t}_i - \Delta t_i$. If $w_i$ reaches its maximum allowed size, the scheduler reduces the number of RBs to allocate for $pkt_i$ to match only the predicted size (i.e., $R_i$ is updated as $R_i = f(\hat{s}_i, mcs_i)$), and searches again for $R_i$ RBs in $w_i$ for $pkt_i$ (lines 8-10 in *Search_in_overlapping_area*). Increasing $w_i$ reduces the likelihood of meeting the latency requirement for packets that deviate significantly from their expected generation time $\hat{t}_i$, but the requirement can still be satisfied in some cases. For example, consider the case where $t_i^{end}$ is equal to $t_i^{end} = \hat{t}_i + L_i$, and the scheduler finds RBs for $pkt_i$ at the end of $w_i$ as illustrated in Fig. 2. In this case, the latency requirement $L_i$ cannot be met if $pkt_i$ is actually generated before $\hat{t}_i$ as illustrated in Fig. 2, but it can still be satisfied if $pkt_i$ is generated after $\hat{t}_i$.

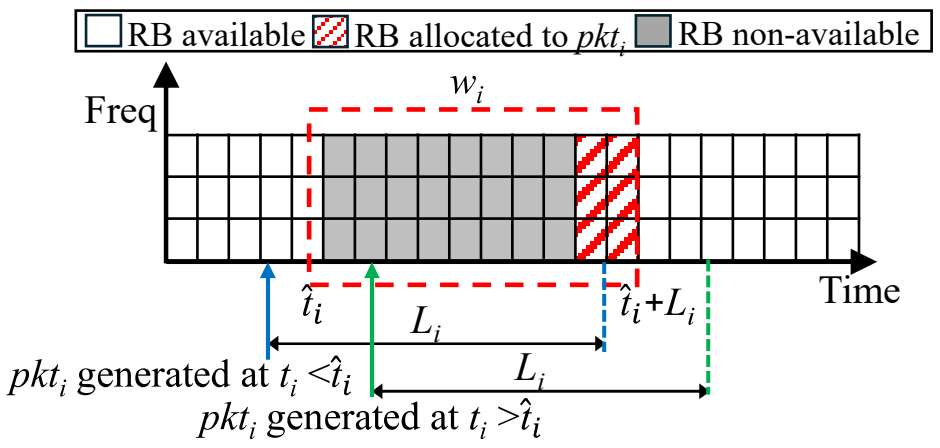

Fig. 2. Example of packet $pkt_i$ generated at $t_i$ for which its latency requirement $L_i$ is satisfied when $t_i < \hat{t}_i$ but not when $t_i < \hat{t}_i$.

## IV. Traffic Characterization and Prediction

The proposed predictive CG scheduler pre-allocates resources for future traffic demands by leveraging predicted traffic information while accounting for potential prediction inaccuracies. To evaluate our proposal, we consider a 6G-envisioned autonomous driving scenario in which sensor data generated by an autonomous vehicle is sent to the network for processing at the edge [10]. Communications must meet a bounded latency deadline to ensure that offloading processing workloads to the network does not disrupt vehicle operations. For our evaluation, we use realistic sensor data generated by autonomous vehicles through a Connected and Automated Mobility (CAM) platform [11]. This platform integrates realistic sensing and autonomous driving (AD) capabilities using the open-source CARLA and AUTOWARE software. We have configured the autonomous vehicle in the CAM platform with a full suite of Level 3 (L3) AD sensors, including five cameras and five radars mounted on the top, front, rear and sides of the vehicle. Each sensor is considered a data source or node in our evaluation, requiring offloading of its traffic to the network. The offloaded traffic includes detected objects such as vehicles, obstacles and pedestrians. Sensors generate raw data at periodic intervals, which is then processed by a perception module to extract detected objects. We have collected extensive datasets of processed sensor packets including their size and timestamp from realistic urban environments (Fig. 3).

The different sensors have a sampling rate of 20 Hz, collecting data every 50 ms. Then, the object detection algorithm introduces a processing delay that depends on the driving scenario and the number of detected objects. We have characterized this processing delay for each sensor, showing a standard deviation ranging from 1.47 to 1.80 ms. Following the nomenclature used in Section III, the proposed predictive CG scheme considers 50 ms as the inter-packet generation time $p$, $\hat{t}_i$ is estimated as $t_i^0 + p$ where $t_i^0$ is the generation time of the first packet for each sensor $i$ of the $N$=10 sensors, and $\Delta t_i$ is set to the characterized standard deviation value $\sigma_i^t$.

The size of processed sensor packets varies significantly, as shown in Fig. 3, since it depends on the number of objects detected in the driving scenario. To predict the size of future processed sensor packets, we implemented a predictor based on a Long Short-Term Memory (LSTM) network using regression-based supervised learning. LSTM networks are particularly suited for tasks involving sequential or time-series data, where the order and context of input elements are relevant. This includes tasks where dependencies between data points over time need to be learned, making it suitable for predicting autonomous vehicle sensor data since the size of each packet is tied to past packets and sensor behaviors. The hyper-parameters used for the configuration of the LSTM are summarized in Table I and were optimized to minimize the Mean Absolute Error (MAE), which quantifies the average magnitude of prediction errors as $\text{MAE}=\sum_{i=1}^{P}|s_i-\hat{s}_i|/P$, where $P$ represents the number of predicted packets, $s_i$ is the actual packet size, and $\hat{s}_i$ is the predicted packet size. We used a dataset containing 9780 samples, with 60% allocated for training the LSTM, 20% for validation, and the remaining 20% for testing. The LSTM network is configured to predict the size of the next ten packets, i.e. the processed sensor packets generated by each of the $N$=10 sensors within the period $p$=50 ms. Fig. 4 shows the CDF (cumulative distribution function) of the absolute error for the prediction of the first, fifth and tenth packets. The figure illustrates the magnitude of possible prediction inaccuracies and reinforces the need to account for such inaccuracies when reserving resources based on predicted traffic demands. As described in Section III, the proposed predicted CG scheduling scheme accounts for the inaccuracy of the predicted packet size $\hat{s}_i$ when determining the resources that should be pre-allocated to each $pkt_i$ to minimize the impact of prediction errors on the capacity to meet the latency requirements.

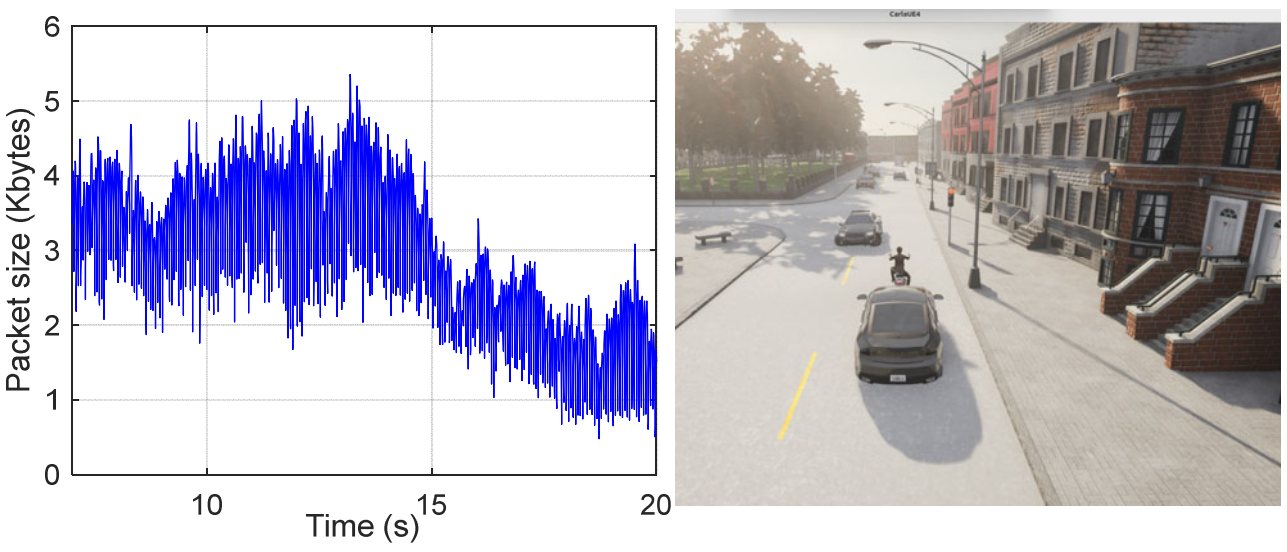

Fig. 3. Sample trace of processed sensor data packets.

TABLE I. LSTM HYPER-PARAMETERS

| Parameter | Value | Parameter | Value |
|---|---|---|---|
| Sequence Length | 150 | Number of layers | 3 (units: 256,128,64) |
| Batch Size | 32 | Features Per Sample | 6 |
| Dropout | 0.1 | Epochs | 100 |
| Optimizer | Adam | Scaling Method | Min-Max Scaling |
| Learning rate | 0.001 | Number of Outputs | 10 |

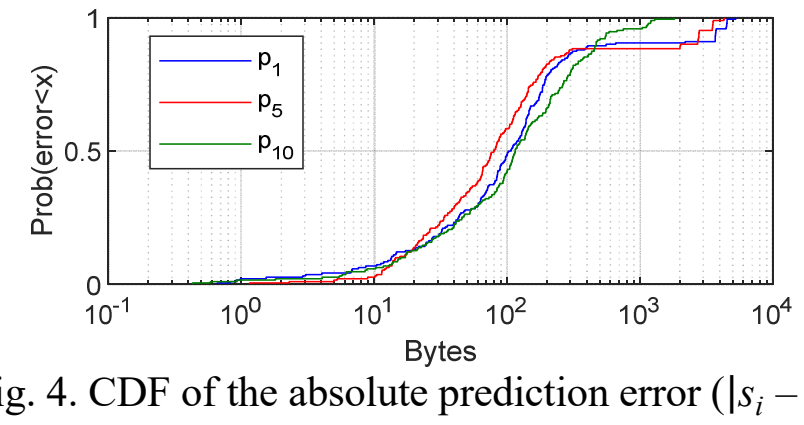

Fig. 4. CDF of the absolute prediction error ($|s_i - \hat{s}_i|$).

## V. EVALUATION AND DISCUSSION

This section compares the performance of the proposed predictive CG scheduler (PCG) against a reference (non-predictive) 5G CG scheduling scheme. The reference scheme periodically allocates RBs for transmitting packets generated by each node. Following [12], the reference scheme allocates RBs to minimize the latency of each packet. For a fair comparison with PCG, the reference scheme uses an allocation period $p$=50 ms and seeks to allocate RBs to $pkt_i$ within $w_i$ = w($t_i^{ini}$, $t_i^{end}$) with $t_i^{ini}=\hat{t}_i+\Delta t_i$, $t_i^{end}=\hat{t}_i-\Delta t_i+L_i$, and $\Delta t_i=\sigma_i^t$, which ensures that a packet generated between $\hat{t}_i$-$\Delta t_i$ and $\hat{t}_i$+$\Delta t_i$ can be transmitted within its latency requirement $L_i$. The reference scheme allocates the amount of RBs necessary to transmit packets of size $S_i$. We test different configurations of the reference scheme where $S_i$ is established based on the most frequent value of the packet sizes, i.e., the mode of the packet size -mode($s_i$)- (C1 configuration), or on the 80th, 90th, 95th, and 99th percentiles of the packet size (C2, C3, C4 and C5 configurations for the reference scheme), respectively. The percentile value is represented by the function $P_x(s_i)$, where $x$ represents the desired percentile. We also evaluate different configurations of the proposed PCG scheduler to assess the impact of the selection of the prediction inaccuracy $\Delta s_i$. PCG is evaluated with values of $\Delta s_i$ equal to the standard deviation of the predicted size $\sigma_i^s$ (C1 configuration for PCG), or equal to the 80th, 90th, 95th, or 99th percentiles of the absolute error of the prediction, i.e., $P_x(AE)$ with $x$=80, 90, 95 and 99, and $AE=|s_i-\hat{s}_i|$ (C2, C3, C4 and C5 configurations for PCG, respectively). Table II shows the evaluated configurations for PCG and the reference schemes. In addition, we also evaluate a predictive CG scheduler that allocates RBs for packet transmissions following a scheduling policy similar to PCG but only considering the predicted packet size $\hat{s}_i$ and not taking into account prediction inaccuracies.

TABLE II. EVALUATED CONFIGURATIONS FOR PCG AND REF.

| Scheduler | Configurations | | | | |
|---|---|---|---|---|---|
| | C1 | C2 | C3 | C4 | C5 |
| PCG | $\Delta s_i = \sigma_i^s$ | $\Delta s_i$=$P_{80}(AE)$ | $\Delta s_i$=$P_{90}(AE)$ | $\Delta s_i$=$P_{95}(AE)$ | $\Delta s_i$=$P_{99}(AE)$ |
| Reference | $S_i$=mode($s_i$) | $S_i$ =$P_{80}(s_i)$ | $S_i$ = $P_{90}(s_i)$ | $S_i$ = $P_{95}(s_i)$ | $S_i$ = $P_{99}(s_i)$ |

The performance of the proposed PCG scheme and the reference scheme is evaluated considering a 5 MHz cell bandwidth, a 30 kHz NR subcarrier spacing (SCS) [4] and a MCS11 (Modulation and Coding Scheme) to balance between robustness and spectral efficiency; similar trends have been observed for other configurations. We consider a mixed traffic flow scenario with latency requirements derived from 3GPP specifications for enhanced V2X scenarios [13]. We consider that 50%, 25% and 25% of the packets must be transmitted within a maximum latency of 50, 20 and 10 ms, respectively; other scenarios with mixed traffic flows and different latency requirements have also been evaluated, and similar trends have been observed.

Fig. 5 presents the percentage of transmissions that meet their latency requirements using the proposed PCG scheme and the reference (Ref.) scheme. The performance is shown for the different configurations evaluated for each scheme following Table II. Fig. 5 shows that PCG significantly increases the percentage of packets that meet their latency requirements compared to the reference scheme. PCG can reach satisfaction levels higher than 87% with all the evaluated configurations, while the reference scheme cannot surpass a 64% satisfaction level regardless of the configuration used. Fig. 5 also reports the performance achieved with a predictive configured grant scheme that does not consider prediction inaccuracies and pre-allocates resources based only on the predicted packet size $\hat{s}_i$. In this case, only about 64% of packets meet their latency requirement (similar to the reference scheme) compared with a satisfaction level of 92.4% and 98.9% with the PCG when $\Delta s_i$=$\sigma_i^s$ (C1) and $\Delta s_i$=$P_{99}(AE)$ (C5) respectively. These results strongly highlight the importance of accounting for prediction inaccuracies in the design of predictive scheduling schemes capable of supporting deterministic communications[1].

[1] We should note that the transmissions that PCG cannot satisfy correspond to packets with a size larger than $\hat{s}_i + \Delta s_i$. In this case, the number of allocated RBs is insufficient for transmitting these packets in time to meet their latency requirement.

Fig. 5 shows that the proposed PCG increases its performance as the prediction inaccuracy $\Delta s_i$ is set equal to higher percentile values of the prediction's absolute error. Similarly, the reference scheme increases its satisfaction level when it allocates RBs based on the highest percentiles of the packet size. However, these improvements come at the expense of pre-allocating a larger number of RBs per packet than actually needed, thus reducing resource efficiency. This is illustrated in Fig. 6, where Fig. 6.a represents the percentage of RBs allocated for the transmission of packets and Fig. 6.b represents the percentage of pre-allocated RBs that are actually unutilized. Fig. 6.a shows that the percentage of allocated RBs increases for PCG when $\Delta s_i$ is set equal to higher percentile values of the prediction's absolute error, and increases for the reference scheme when allocating RBs based on the highest size percentiles. However, PCG pre-allocates in general a lowest number of RBs than the reference scheme while achieving higher satisfaction levels (Fig. 5). In addition, PCG results in a significantly lower percentage of non-utilized RBs compared to the reference scheme, which highlights its highest resource efficiency.

Following the URLLC principles in 5G, the reference scheme focuses on minimizing the transmission latency. On the other hand, PCG has been designed to support deterministic communications in beyond 5G networks, and hence prioritizes maximizing the number of transmissions that meet their latency requirements. Fig. 7 plots the cumulative distribution function (CDF) of the latency experienced by packets with 10, 20, and 50 ms latency requirements, respectively. The figure shows that packets with the most stringent latency requirement (10 ms) experience the lowest latency values when using the PCG scheme. On the other hand, PCG slightly increases the latency of packets with more relaxed latency constraints. This is particularly the case for packets with a 20 ms requirement as these packets often compete for the same RBs as the packets that require 10 ms latency. PCG intentionally delays the transmission of packets with more relaxed latency requirements, while still meeting their latency deadline, to ensure that packets with tighter deadlines can be transmitted in time. In contrast, a scheduler that seeks minimizing the transmission latency (Ref.) does not leverage varying latency requirements in mixed traffic flows to increase the percentage of satisfied transmissions, and the latency experienced increases uniformly up to the maximum allowed latency.

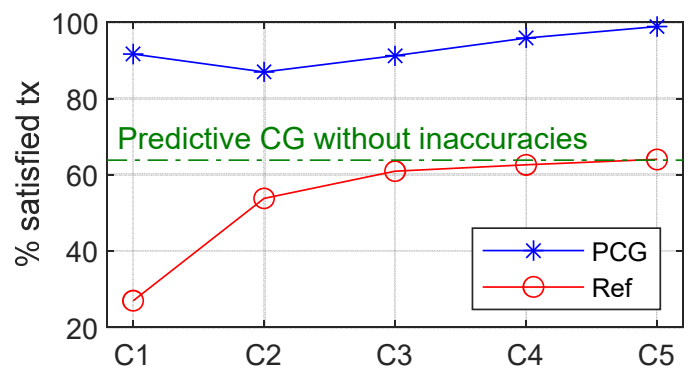

Fig. 5. Percentage (%) of transmissions meeting their latency requirements.

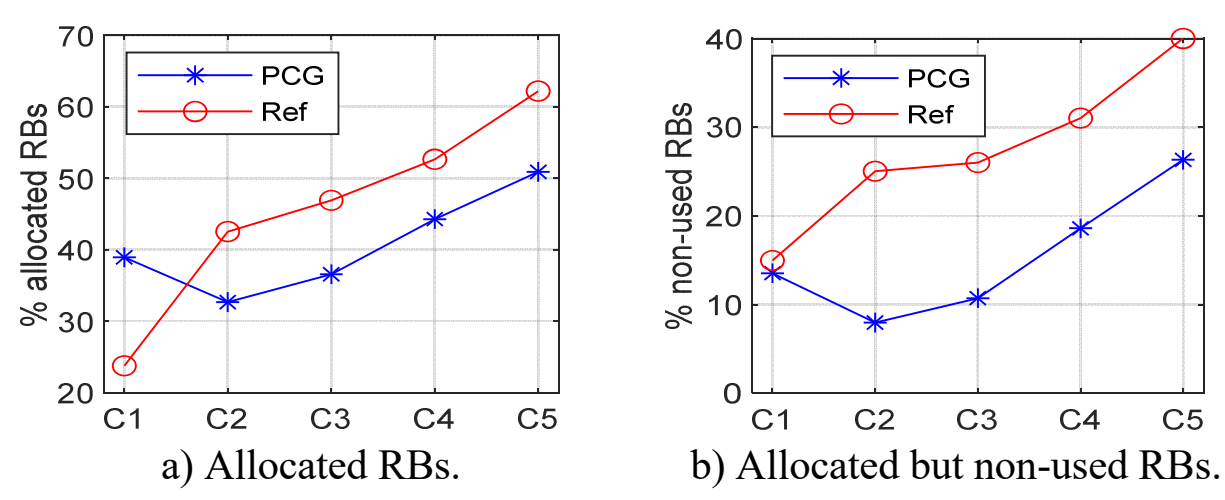

a) Allocated RBs. b) Allocated but non-used RBs.

Fig. 6. Percentage of allocated RBs and percentage of allocated RBs that are non-used for packet transmissions.

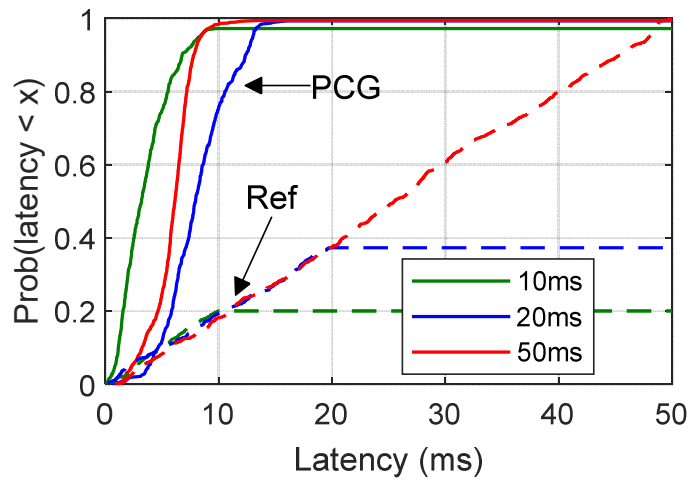

Fig. 7. CDF of the latency experienced by packets with 10, 20 and 50 ms latency requirements with PCG and the reference scheme.

## VI. Conclusions

This study presents a novel predictive configured grant scheduling scheme designed to support deterministic communications in beyond 5G networks. The proposed predictive CG scheduling scheme pre-allocates resources based on traffic predictions and prediction inaccuracies with the objective to maximize the percentage of packets transmitted within their bounded latency requirements. The evaluation shows that the proposed predictive CG scheme significantly increases the percentage of satisfied transmissions compared to a 5G CG scheduler while pre-allocating less resources and using them more efficiently. Our evaluation has also demonstrated the importance of accounting for prediction inaccuracies in the scheduling process to improve the ability to meet bounded latency requirements and hence support deterministic services in beyond 5G networks.